\documentstyle{article}

\begin{document}

\begin{center}
PHOTOMETRY AND PHOTOMETRIC REDSHIFTS OF GALAXIES IN THE HUBBLE DEEP FIELD SOUTH
NICMOS FIELD
\end{center}

  We present an electronic catalog of infrared and optical photometry and
photometric redshifts of 323 galaxies in the Hubble Deep Field South NICMOS
field at http://www.ess.sunysb.edu/astro/hdfs/home.html. The analysis is based
on infrared images obtained with the Hubble Space Telescope using the Near
Infrared Camera and Multi-Object Spectrograph and the Space Telescope Imaging
Spectrograph together with optical images obtained with the Very Large
Telescope.  The infrared and optical photometry is measured by means of a new
quasi-optimal photometric technique that fits model spatial profiles of the
galaxies determined by Pixon image reconstruction techniques to the images.  In
comparison with conventional methods, the new technique provides higher
signal-to-noise-ratio measurements and accounts for uncertainty correlations
between nearby, overlapping neighbors.  The photometric redshifts are measured
by means of our redshift likelihood technique, incorporating six
spectrophotometric templates which, by comparison with spectroscopic redshifts
of galaxies identified in the Hubble Deep Field North, are known to provide
redshift measurements accurate to within an RMS relative uncertainty of $\Delta
z/(1 + z) < 0.1$ at all redshifts $z < 6$.  The analysis reaches a peak
$H$-band sensitivity threshold of $AB(16000) = 28.3$ and covers 1.02 arcmin$^2$
to $AB(16000) = 27$, 1.27 arcmin$^2$ to $AB(16000) = 26$, and 1.44 arcmin$^2$
to $AB(16000) = 25$.  The analysis identifies galaxies at redshifts ranging
from $z \approx 0$ through $z > 10$, including 17 galaxies of redshift $5 < z <
10$ and five candidate galaxies of redshift $z > 10$.  (The redshift likelihood
functions are given, allowing high-redshift galaxies with sharply peaked
redshift likelihood functions to be distinguished from candidate high-redshift
galaxies with broad or bimodal redshift likelihood functions.)  The analysis
can also be used to establish firm upper limits to the surface densities of
galaxies as a function of brightness and redshift to redshifts as large as $z =
14$.

\end{document}